\documentclass{article}                  
\usepackage{amsmath} 
\usepackage{amsthm} 
\usepackage{latexsym} 
\usepackage{amssymb}
\usepackage{color}
\usepackage{graphics}

\newtheorem{theorem}{Theorem}
\newtheorem{lemma}{Lemma}
\newtheorem{remark}{Remark}

\theoremstyle{definition}
\newtheorem{definition}{Definition}
\newtheorem{notation}{Notation}

\newcommand{\fA}{\ensuremath{\mathfrak{A}}} 
\newcommand{\fB}{\ensuremath{\mathfrak{B}}} 
\newcommand{\N}{\ensuremath{\mathbb{N}}}   
\newcommand{\cC}{\ensuremath{\mathcal{C}}}   
\newcommand{\cGC}{\ensuremath{\mathcal{GC}}}   
\newcommand{\cE}{\ensuremath{\mathcal{E}}}   

\newcommand{\myvec}[1]{\ensuremath{\mathbf{#1}}}   

\begin{document}
\title{Complexity of the Guarded Two-Variable Fragment with
Counting Quantifiers}
\author{Ian Pratt-Hartmann\\
School of Computer Science\\
Manchester University}
\date{}
\maketitle
\begin{abstract}
  We show that the finite satisfiability problem for the guarded
  two-variable fragment with counting quantifiers is in EXPTIME. The
  method employed also yields a simple proof of the result obtained in
  Kazakov~\cite{logic:kazakov04}, that the satisfiability problem for
  the guarded two-variable fragment with counting quantifiers is in
  EXPTIME.
\end{abstract}
\section{Introduction}
\label{sec:intro}
The {\em two-variable fragment with counting quantifiers}, here
denoted $\cC^2$, is the set of function-free, first-order formulas
containing at most two variables, but with the counting quantifiers
$\exists_{\leq C}$, $\exists_{\geq C}$ and $\exists_{= C}$ (for every
$C >0$) allowed. The {\em guarded two-variable fragment with counting
quantifiers}, here denoted $\cGC^2$, is the subset of $\cC^2$ whose
formulas contain no individual constants, and where all quantifiers
appear only in guarded patterns (explained below). Both $\cC^2$ and
$\cGC^2$ are assumed to contain equality. Neither fragment has the
finite model property, so that their respective satisfiability
problems and finite satisfiability problems do not coincide.  It was
shown in Pratt-Hartmann~\cite{logic:ph05} that the satisfiability and
finite satisfiability problems for $\cC^2$ are both in NEXPTIME.  It
was shown in Kazakov~\cite{logic:kazakov04} that the satisfiability
problem for $\cGC^2$ is in EXPTIME; however the method employed in
that paper yields no information about the finite satisfiability
problem for $\cGC^2$.  In this paper, we show that the finite
satisfiability problem for $\cGC^2$ is in EXPTIME; furthermore, the
method employed here also yields a simple proof of Kazakov's result.
All complexity results mentioned in this paper assume succinct
(binary) coding of numerical quantifier subscripts.

The fragment $\cGC^2$ is a proper superset of the ``description
logic'' $\mathcal{ALCQI}$ (Calvanese~\cite{logic:calvanese96}), which
in turn is a proper superset of the description logic $\mathcal{ALC}$
with general concept inclusion (essentially, multimodal $K$ with
universal quantification of formulas). This latter fragment 
has the finite model property, and its satisfiability problem (=
finite satisfiability problem) is EXPTIME-hard. Hence, the complexity
bounds reported here for $\cGC^2$ are tight.  Furthermore, it was
shown in Lutz {\em et al.}~\cite{logic:lst03,logic:lst05} that
the finite satisfiability problem for $\mathcal{ALCQI}$ is in
EXPTIME. The present paper thus extends that result to the
whole of $\cGC^2$.

By contrast, $\cGC^2$ is a proper subset of $\cC^2$, whose
satisfiability and finite satisfiability problems are
NEXPTIME-hard. In fact, even the 2-variable fragment of first-order
logic {\em without} counting quantifiers, which has the finite model
property, exhibits a NEXPTIME-hard satisfiability problem, as can be
shown by using its formulas to encode exponentially large grids (see,
e.g.~B\"{o}rger {\em et al.}~\cite{logic:BGG97},
pp.~253~ff.). In such encodings, the quantifiers ($\forall$ and
$\exists$) appear in unguarded patterns. On the other hand, in the
presence of counting quantifiers, a similar encoding is possible using
only guarded quantification, provided that just one individual
constant is admitted to the language. Thus, both the lack of
individual constants and the restriction to guarded patterns of
quantification are essential to the comparatively low complexity of
$\cGC^2$ (assuming, of course, that EXPTIME $\neq$ NEXPTIME).

It was shown in Pratt-Hartmann~\cite{logic:ph05}, Corollary~1 that,
if a formula $\phi$ of $\cC^2$ is finitely satisfiable, then it has a
model whose size is bounded by a doubly exponential function of the
number of symbols in $\phi$. This bound is optimal, even for
the fragment $\cGC^2$, in the sense that there exists a sequence
$\{\phi_i\}$ of finitely satisfiable formulas of $\cGC^2 \subseteq
\cC^2$ whose size grows as a polynomial function of $i$, but whose
smallest satisfying structures grow as a doubly exponential function
of $i$ (Gr\"{a}del {\em et al.}~\cite{logic:gor97}, p.~317). In view of
this lower bound on the sizes of smallest satisfying structures, the
upper complexity bound on the finite satisfiability problem for
$\cGC^2$ proved here is noteworthy.

The plan of this paper is as follows. Section~\ref{sec:prelim} defines
the fragment $\cGC^2$ and establishes some basic results.
Section~\ref{sec:trans} describes a procedure for transforming any
$\cGC^2$-formula into an (exponentially large) constraint
satisfaction problem with variables ranging over $\N$.
Section~\ref{sec:main} uses this transformation to prove that the
finite satisfiability problem for $\cGC^2$ is in EXPTIME.
Section~\ref{sec:dessert} is dessert.

\section{Preliminaries}
\label{sec:prelim}
We restrict consideration to first-order languages whose only
primitive symbols are the variables $x$ and $y$, the usual Boolean
connectives, the quantifiers $\forall$, $\exists$, $\exists_{\leq C}$,
$\exists_{\geq C}$, $\exists_{= C}$ (for all $C > 0$), the equality
predicate (here written as $\approx$) and a non-logical signature of
0-ary, unary and binary predicates. There are no individual constants
or function-symbols in these languages.  If $p$ is any binary
predicate (including $\approx$), we call an atomic formula having
either of the forms $p(x,y)$ or $p(y,x)$ a {\em guard-atom}. The
two-variable guarded fragment with counting quantifiers, $\cGC^2$, can
then be defined as the smallest set of formulas satisfying the
following conditions:
\begin{enumerate}
\item $\cGC^2$ contains all atomic formulas and is closed under
Boolean combinations;
\item if $\phi$ is a formula of $\cGC^2$ with at most one free
variable, and $u$ is a variable (i.e.~either $x$ or $y$), then the
formulas $\forall u \phi$ and $\exists u \phi$ are in $\cGC^2$;
\item if $\phi$ is a formula of $\cGC^2$, $\gamma$ a guard-atom,
$u$ a variable, and $Q$ any of the quantifiers $\exists$,
$\exists_{\leq C}$, $\exists_{\geq C}$, $\exists_{= C}$ (for 
$C >0$), then the
formulas $\forall u (\gamma \rightarrow \phi)$ and $Q u (\gamma \wedge
\phi)$ are in $\cGC^2$.
\end{enumerate}
We take the semantics to be the usual semantics of first-order logic,
with counting quantifiers interpreted in the obvious way.  If $\phi$
is any formula of $\cGC^2$, we denote the number of symbols in $\phi$
by $\lVert \phi \rVert$, assuming binary coding of numerical
quantifier subscripts.

According to the above syntax, the non-counting quantifiers $\exists$
and $\forall$ may apply without restriction to formulas with at most
one free variable; however, they may apply to formulas with two free
variables only in the presence of a guard-atom.  By contrast, the
counting quantifiers $\exists_{\leq C}$, $\exists_{\geq C}$,
$\exists_{= C}$ may only every apply in the presence of a guard atom
(which by definition has two free variables). Note in particular that
the formula $\exists_{=1} x p(x)$ is not in $\cGC^2$.  In fact, the
next lemma shows that no formula of $\cGC^2$ can force a predicate $p$
to be uniquely instantiated in its models.
\begin{definition}
Let $\fA$, $\fB$ be structures over disjoint domains $A$, $B$,
 respectively, interpreting a common signature $\Sigma$ with no
 individual constants. The {\em union} $\fA \cup \fB$ of
 $\fA$ and $\fB$ is the structure with domain $A \cup B$ and
 interpretations $\sigma^{\fA \cup \fB} = \sigma^{\fA} \cup
 \sigma^{\fB}$ for every $\sigma \in \Sigma$.
\label{def:dup} 
\end{definition}
\begin{lemma}
Let $\phi$ be a formula of $\cGC^2$ and $\fA$ a structure over the 
signature of $\phi$. For $N \geq 1$, let $N \cdot \fA$ denote the union of $N$
disjoint copies of $\fA$. If $\phi$ is satisfied in $\fA$, then it is satisfied
in $N \cdot \fA$.
\label{lma:dup}
\end{lemma}
\begin{proof}
If $\theta: \{x,y\} \rightarrow A$ is any variable assignment over $A$
and $1 \leq i \leq N$, let $\theta^i$ be the variable assignment over
$N \cdot \fA$ which maps $x$ and $y$ to the corresponding elements in
the $i$th copy of $\fA$. A routine structural induction on $\phi$
shows that $\fA \models_\theta \phi$ if and only if, for some (= for
all) $i$ ($1 \leq i \leq N$), $N \cdot \fA \models_{\theta^i} \phi$.
\end{proof}
It follows immediately from Lemma~\ref{lma:dup} that, if a formula of
$\cGC^2$ has a finite model, then it has arbitrarily large finite
models, and indeed infinite models.

The following lemma, which is a cosmetic modification of
Kazakov~\cite{logic:kazakov04}, \linebreak
Lemma~2, establishes a normal form for
$\cGC^2$-formulas.
\begin{lemma}
Let $\psi$ be a formula in $\cGC^2$. We can construct, in time bounded
by a polynomial function of $\lVert \psi \rVert$, a formula
\begin{equation}
\begin{split}
\phi := 
\forall x \alpha \wedge 
 \bigwedge_{1 \leq h \leq l} \forall x & \forall y  
     (e_h(x,y) \rightarrow (\beta_h \vee x \approx y)) \wedge \\
  &  \bigwedge_{1 \leq i \leq m} 
      \forall x \exists_{=C_i} y (f_i(x,y) \wedge x \not \approx y)
\end{split}
\label{eq:normalform}
\end{equation}
such that: $(i)$ $\alpha$ is a quantifier-free formula not involving
$\approx$ with $x$ as its only variable; $(ii)$ $l$ and $m$ are
positive integers; $(iii)$ for all $h$ $(1 \leq h \leq l)$, $e_h$ is a
binary predicate other than $\approx$, and $\beta_h$ is a
quantifier-free formula not involving $\approx$ with $x$ and $y$ as
its only variables; $(iv)$ for all $i$ $(1 \leq i \leq m)$, $C_i$ is a
positive integer, and $f_i$ is a binary predicate other than $\approx$;
$(v)$ $\phi$ is satisfiable if and only if $\psi$ is satisfiable;
$(vi)$ $\phi$ is finitely satisfiable if and only if $\psi$ is
finitely satisfiable.
\label{lma:normalform}
\end{lemma}
\begin{proof}
Standard transformation to Scott normal form, using
Lemma~\ref{lma:dup}. See Kazakov {\em op.~cit.}~for details.
\end{proof}
Hence, to show that the (finite) satisfiability problem for $\cGC^2$
is in EXPTIME, it suffices to consider only formulas of the
form~\eqref{eq:normalform}.  Furthermore, we may assume without loss
of generality that no 0-ary predicates (proposition letters) occur in
$\phi$, since we can consider each of the (at most $2^{\lVert \phi
\rVert}$) truth-value assignments to the 0-ary predicates of $\phi$ in turn,
replacing each 0-ary predicate with $\top$ or $\bot$ according to its
truth-value in the considered assignment.

Accordingly, fix $\phi$ to be some formula of the
form~\eqref{eq:normalform} over a signature of unary and binary
predicates. Set $C = \max_{1 \leq i \leq m} C_i$, and let $\Sigma$ be
the signature of $\phi$ together with $\log((mC)^2 +1)$ (rounded up)
new unary predicates. Thus, $|\Sigma|$ is bounded by a polynomial
(actually, linear) function of $\lVert \phi \rVert$.  Since $\Sigma$
is the only signature we shall be concerned with in the sequel, we
generally suppress reference to it.  Thus, `predicate' henceforth
means `predicate in $\Sigma \cup \{ \approx \}$', `structure' henceforth
means `structure interpreting $\Sigma$', and so on.  We keep the
meanings of the symbols $\alpha$, $l$, $m$, $e_h$, $\beta_h$, ($1
\leq h \leq l$), $C_i$, $f_i$ ($1 \leq i \leq l$), $\phi$, $C$,
$\Sigma$ fixed throughout this paper. The predicates $f_1, \ldots,
f_m$ will play a key role in the ensuing argument; we refer to them as
the {\em counting predicates}. There is no restriction on these
predicates' occurring in other parts of $\phi$: in particular, they
may feature as guards.

We review some standard concepts.  A {\em literal} is an atomic
formula or the negation of an atomic formula.  A {\em 1-type} is a
maximal consistent set of equality-free literals involving only the
variable $x$. A {\em 2-type} is a maximal consistent set of
equality-free literals involving only the variables $x$ and $y$. If
$\tau$ is a 2-type, then the result of transposing the variables $x$
and $y$ in $\tau$ will also be a 2-type, denoted $\tau^{-1}$.  If
$\fA$ is any structure, and $a \in A$, then there exists a unique
1-type $\pi(x)$ such that $\fA \models \pi[a]$; we denote $\pi$ by
${\rm tp}^\fA[a]$. If, in addition, $b \in A$ is distinct from $a$,
then there exists a unique 2-type $\tau(x,y)$ such that $\fA \models
\tau[a,b]$; we denote $\tau$ by ${\rm tp}^\fA[a,b]$.  We do not define
${\rm tp}^\fA[a,b]$ if $a = b$.
\begin{notation} 
  Any 2-type $\tau$ includes a unique 1-type, denoted ${\rm
  tp}_1(\tau)$; in addition, we write ${\rm tp}_2(\tau)$ for ${\rm
  tp}_1(\tau^{-1})$.
\label{notation:types}
\end{notation}
\begin{remark}
Let $\fA$ be a structure, and let $a$, $b$ be distinct elements of
$A$. If \/ ${\rm tp}^\fA[a,b] = \tau$, then ${\rm
tp}^\fA[b,a] = \tau^{-1}$, ${\rm tp}^\fA[a] = {\rm tp}_1(\tau)$, and
${\rm tp}^\fA[b] = {\rm tp}_2(\tau)$.
\label{remark:types}
\end{remark}
\begin{definition}
  Let $\tau$ be a 2-type.  We say that $\tau$ is a {\em message-type}
  if, for some counting predicate $f_i$ ($1 \leq i \leq m$), $f_i(x,y)
  \in \tau$.  If $\tau$ is a message-type such that $\tau^{-1}$ is
  also a message-type, we say that $\tau$ is {\em invertible};
  otherwise, $\tau$ is {\em non-invertible}. Finally, if $\tau$ is a
  2-type such that neither $\tau$ nor $\tau^{-1}$ is a message-type,
  we say that $\tau$ is {\em silent}.
\label{def:messagetype}
\end{definition}
\begin{remark}
Let $\fA$ be a structure, and let $a$, $b$ be distinct elements of
$A$. Then ${\rm tp}^\fA[a,b]$ is a message-type just in case $\fA
\models f_i[a,b]$ for some $i$ $(1 \leq i \leq m)$; if so, then this
message-type is invertible just in case $\fA \models f_{i'}[b,a]$ for
some $i'$ $(1 \leq i' \leq m)$.
\label{remark:messagetypes}
\end{remark}
The terminology is meant to suggest the following imagery.  If ${\rm
tp}^\fA[a,b]$ is a message-type $\mu$, then we may imagine that $a$
sends a message (of type $\mu$) to $b$. If $\mu$ is invertible, then
$b$ replies by sending a message (of type $\mu^{-1}$) back to $a$.  If
${\rm tp}^\fA[a,b]$ is silent, then neither element sends a message to
the other.

\begin{definition}
  Let $\fA$ be a structure.  We say that $\fA$ is {\em chromatic} if
  distinct elements connected by a chain of 1 or 2 invertible
  message-types have distinct 1-types. That is, $\fA$ is chromatic
  just in case, for all $a, a', a'' \in A$:
\begin{enumerate}
\item if $a \neq a'$ and ${\rm tp}^\fA[a,a']$ is an invertible
  message-type, then ${\rm tp}^\fA[a] \neq {\rm tp}^\fA[a']$; and
\item if $a, a', a''$ are pairwise distinct and both ${\rm
  tp}^\fA[a,a']$ and ${\rm tp}^\fA[a',a'']$ are invertible
  message-types, then ${\rm tp}^\fA[a] \neq {\rm tp}^\fA[a'']$.
\end{enumerate}
\label{def:chromatic}
\end{definition}
\begin{lemma}
  If $\phi$ has a model, then it has a chromatic model over the same
  domain.
\label{lma:chromatic}
\end{lemma}
\begin{proof}
  Suppose $\fA \models \phi$, and consider the (undirected) graph $G$
  on $A$ whose edges are the pairs of distinct elements connected by a
  chain of 1 or 2 invertible message-types. That is, $G = (A,E^1 \cup
  E^2)$, where
\begin{center}
\begin{minipage}{20cm}
\begin{tabbing}
$E^1 = $\= $\{ (a,a') \mid$\= $\mbox{$a \neq a'$ and ${\rm tp}^{\fA}[a,a']$ is an 
                           invertible message-type} \}$ \\
$E^2 = $\> $\{ (a,a'') \mid \mbox{$a \neq a''$ and for some $a' \in A$,
                            $(a,a')$ and $(a',a'')$ are}$ \\ 
\> \>  $\mbox{ both in $E^1$}\}$. 
\end{tabbing}
\end{minipage}
\end{center}
Since $C = \max_{1 \leq i \leq m} C_i$, the degree of $G$ (in the
normal graph-theoretic sense) is at most $(mC)^2$. Now use the
standard (greedy) algorithm to colour the nodes of $G$ with $(mC)^2 +
1$ colours in such a way that no edge joins two nodes of the same
colour.  By interpreting the $\log((mC)^2 + 1)$ (rounded up) unary
predicates of $\Sigma$ not occurring in $\phi$ to encode these
colours, we obtain the desired chromatic model.
\end{proof}

In the sequel, we shall need to refer to sets of invertible
message-types indexed by bit-strings as follows.  Let the 1-types be
enumerated as
\begin{equation*}
\Pi = \pi_0, \ldots, \pi_{P-1}.
\end{equation*}
Evidently, $P$ is a power of 2, so $p = \log P$ is an
 integer. (Actually, $p = |\Sigma|$.) Now
let $s$ be any bit string ($0 \leq |s| \leq p$), and denote the string of
length 0 by $\epsilon$. We inductively define the sub-sequence
$\Pi_s$ of $\Pi$ by setting $\Pi_\epsilon$ to be the whole of $\Pi$,
and setting $\Pi_{s0}$ and $\Pi_{s1}$ to be the left and right halves
of $\Pi_{s}$, respectively. Formally:
\begin{eqnarray*}
\Pi_\epsilon & = & \pi_0, \ldots, \pi_{P-1},
\end{eqnarray*}
and if $\Pi_{s} = \pi_j, \ldots, \pi_{k-1}$, with $|s| < p$,
\begin{eqnarray*}
\Pi_{s0} & = & \pi_j, \ldots, \pi_{\frac{k - j}{2}} \\
\Pi_{s1} & = &  \pi_{\frac{k - j}{2} +1}, \ldots, \pi_{k-1}.
\end{eqnarray*}
Thus, if $|s| = p$, then $\Pi_s$ is a one-element sequence $\pi_j$,
where $j$ is the integer ($0 \leq j < P$) encoded by the bit-string
$s$ in the usual way. To avoid clumsy circumlocutions, we occasionally
equivocate between bit-strings and the integers they encode, thus, for
example, writing $\pi_s$ instead of $\pi_j$ in this case. But we will
only ever write $\pi_s$ if $|s| = p$.

We may use the sets $\Pi_s$ to define sets of invertible message-types
indexed by bit-strings as follows. Fix any 1-type $\pi$, and denote by
$\Lambda_\pi$ the set of invertible message-types $\lambda$ such that
${\rm tp}_1(\lambda) = \pi$.  If $s$ is any bit-string such that $|s|
\leq p$, let
\begin{equation*}
\Lambda_{\pi,s} = 
   \{\lambda \in \Lambda_\pi \mid {\rm tp}_2(\lambda) \in \Pi_s \}.
\end{equation*}
Thus, the $\Lambda_{\pi,s}$ are sets of invertible message-types
identified purely by their terminal 1-types. Except in very special
cases, these sets will contain more than one member, even when $|s| =
p$. However, for chromatic models, we have the following important
fact.
\begin{lemma}
If $\fA$ is chromatic, $\pi = {\rm tp}^\fA[a]$, and $s$ is a bit-string
with $|s| = p$, then there can be at most one element $b \in A
\setminus \{a\}$ 
such that ${\rm tp}^\fA[a,b] \in \Lambda_{\pi,s}$. 
\label{lma:onlyone}
\end{lemma}
\begin{proof}
Any two such elements would be connected by a chain of two invertible
message-types, and would both have the 1-type $\pi_s$.
\end{proof}

Finally, we use bit strings to index other sets of 2-types as follows.
Again, fix any 1-type $\pi$, and consider the set of non-invertible
message-types $\mu$ such that ${\rm tp}_1(\mu) = \pi$.  Let these be
enumerated in some way as a sequence
\begin{equation*}
\mu_{\pi,0}, \ldots, \mu_{\pi, R-1}.
\end{equation*}
Furthermore, consider the set of silent 2-types $\mu$ such that ${\rm
tp}_1(\mu) = \pi$.  Let these be enumerated in some way as a sequence
\begin{equation*}
\mu_{\pi,R}, \ldots, \mu_{\pi, Q-1}.
\end{equation*}
Thus, the sequence
\begin{equation*}
M_\pi = \mu_{\pi,0}, \ldots, \mu_{\pi, Q-1}.
\end{equation*}
is a list of precisely those 2-types $\tau$ such that ${\rm
tp}_1(\tau) = \pi$ and $\tau^{-1}$ is not a message-type.  Evidently,
$R$ and $Q$ are independent of the choice of $\pi$; moreover, $Q$ is a power of 2,
so $q = \log Q$ is an integer.  (We remark that $R$ is not a power of
2.)  Let $t$ be any bit string ($0 \leq |t| \leq q$). We inductively
define the sub-sequence $M_{\pi,t}$ of $M_\pi$ by setting $M_{\pi,
\epsilon}$ to be the whole of $M_\pi$, and setting $M_{\pi, t0}$ and
$M_{\pi, t1}$ to be the left and right halves of $M_{\pi, t}$,
respectively. Formally:
\begin{eqnarray*}
M_{\pi, \epsilon} & = & \mu_{\pi,0}, \ldots, \mu_{\pi, Q-1},
\end{eqnarray*}
and if $M_{\pi, t} = \mu_{\pi, j}, \ldots, \mu_{\pi, k-1}$, with $|t| < q$,
\begin{eqnarray*}
M_{\pi, t0} & = & \mu_{\pi,j}, \ldots, \mu_{\pi, \frac{k - j}{2}}\\
M_{\pi, t1} & = & \mu_{\pi, \frac{k - j}{2} +1}, \ldots, \mu_{\pi, k-1}. 
\end{eqnarray*}
Thus, if $|t| = q$, then $M_{\pi, t}$ is a one-element sequence
$\mu_{\pi,j}$, where $j$ is the integer ($0 \leq j < Q$) encoded by
the bit-string $t$ in the usual way. Again we may for convenience
write $\mu_{\pi,t}$ instead of $\mu_{\pi,j}$ in this case, but here
too we only ever write $\mu_{\pi,t}$ if $|t| = q$. 

\section{Transformation into an integer constraint\\  problem}
\label{sec:trans}
Henceforth, {\em vector} means ``$m$-dimensional vector over $\N$\/''.
If $\myvec{u}$ and $\myvec{v}$ are vectors, we write $\myvec{u} \leq
\myvec{v}$ if every component of $\myvec{u}$ is less than or equal to
the corresponding component of $\myvec{v}$; we write $\myvec{u} <
\myvec{v}$ if $\myvec{u} \leq \myvec{v}$ and $\myvec{u} \neq
\myvec{v}$. Similarly for $\geq$ and $>$.  The number of vectors
$\myvec{u}$ such that $\myvec{u} \leq \myvec{C}$ is bounded by
$(C+1)^m$, and hence by an exponential function of $\lVert \phi
\rVert$.

Referring to the formula~\eqref{eq:normalform}, denote the vector
$(C_1, \ldots, C_m)$ by $\myvec{C}$ and the vector $(0, \ldots, 0)$ by
$\myvec{0}$.  Moreover, given any 2-type $\tau$, we write
$\myvec{C}_\tau$ for the vector $(C_{\tau, 1}, \ldots, C_{\tau, m})$
where, for all $i$ ($1 \leq i \leq m$),
\begin{equation}
C_{\tau,i} =
\begin{cases}
1 \text{ \qquad if $f_i(x,y) \in \tau$, } \\
0 \text{ \qquad otherwise.}
\end{cases}
\label{eq:Ctau}
\end{equation}
Note that, if $\tau$ is not a message-type---in particular, if $\tau$
is a silent 2-type---we have $\myvec{C}_\tau = \myvec{0}$.  

Let $\tau$ be any 2-type. Since $\tau$ is a finite set of formulas
with free variables $x$ and $y$, we may write $\bigwedge \tau$ to
denote their conjunction.  Referring again to the
formula~\eqref{eq:normalform}, we say that $\tau$ is {\em forbidden},
if the formula
\begin{equation}
\alpha(x) \wedge 
\alpha(y) \wedge 
 \bigwedge_{1 \leq h \leq l} 
     (e_h(x,y) \rightarrow \beta_h) \wedge 
   \bigwedge \tau
\label{eq:forbidden}
\end{equation}
is unsatisfiable. Thus, if $\fA \models \phi$ and $a$, $b$ are
distinct elements of $A$, then ${\rm tp}^\fA[a,b]$ cannot be forbidden.
Since~\eqref{eq:forbidden} is purely Boolean, we can evidently
identify the forbidden 2-types in time bounded by an exponential
function of $\lVert \phi \rVert$.

In the sequel, we take $\pi$ to vary over the set of 1-types,
$\lambda$ to vary over the set of invertible message-types, $s$ to
vary over the set of bit-strings of length at most $p$, $t$ to vary
over the set of bit-strings of length at most $q$, and $\myvec{u}$,
$\myvec{v}$ and $\myvec{w}$ to vary over the set of vectors $\leq
\myvec{C}$. (Similarly for their primed counterparts $\pi'$,
$\lambda'$, $s'$, $t'$, $\myvec{u'}$, $\myvec{v'}$ and $\myvec{w'}$.)
We refer to these sets as the {\em standard ranges} of the respective
letters.  Occasionally, additional restrictions on these ranges will
be imposed.

Now let $V$ be the set whose elements are the following (distinct)
symbols, where the indices $\lambda$, $\pi$, $s$, $t$,
$\myvec{u}$, $\myvec{v}$, $\myvec{w}$ vary over their standard ranges:
\begin{equation*}
\begin{array}{lll}
x_\lambda, \hspace{0.5cm} & y_{\pi,s,\myvec{u}}, & z_{\pi,t,\myvec{u}}, \\ 
\ & \hat{y}_{\pi,s,\myvec{v},\myvec{w}}
    \text{ whenever $|s| < p$}, \hspace{0.5cm} &
\hat{z}_{\pi,t,\myvec{v},\myvec{w}}
    \text{ whenever $|t| < q$}.
\end{array}
\end{equation*}
\noindent
The symbols $\hat{y}_{\pi,s,\myvec{v},\myvec{w}}$ and
$\hat{z}_{\pi,t,\myvec{v},\myvec{w}}$ are not defined when $|s| = p$
and $|t| = q$.  The cardinality of $V$ is evidently bounded by an
exponential function of $\lVert \phi \rVert$.  We impose some
arbitrary order on $V$, and refer to its elements as {\em variables}.
If $U$ is a non-empty set of variables, enumerated, in order, as
$\{u_1, \ldots, u_k\}$, let $\sum U$ denote the term $u_1 + \cdots
+u_k$; if $U$ is the empty set, let $\sum U$ denote the (constant)
term $0$. In the sequel, we take a {\em constraint} to be an equation
or inequality involving arithmetical terms over $V$, or a conditional
statement formed from two such inequalities. A {\em solution} of a set
of constraints over some numerical domain $\mathbb{D}$ is simply a
function $\theta: V \rightarrow \mathbb{D}$ under which all the
constraints in question evaluate (in the obvious way) to true.

With this notation, let $\cE_1$ be the following set of constraints
involving the variables $V$, where $\pi$, $\myvec{u}$, $\myvec{v}$,
$\myvec{w}$ again vary over their standard ranges, and $s$, $t$ vary
over bit-strings such that $|s| <p$ and $|t| < q$:
\begin{eqnarray}
z_{\pi,\epsilon,\myvec{u}} & = & 
y_{\pi,\epsilon,\myvec{C} -\myvec{u}} 
\label{eq:2} \\
y_{\pi,s,\myvec{u}} & = & 
\sum \{ \hat{y}_{\pi,s,\myvec{v}',\myvec{w}'} \mid 
         \myvec{v}' + \myvec{w}' = \myvec{u} \}
\label{eq:3} \\
z_{\pi,t,\myvec{u}} & = & 
\sum \{ \hat{z}_{\pi,t,\myvec{v}',\myvec{w}'} \mid 
         \myvec{v}' + \myvec{w}' = \myvec{u} \}
\label{eq:6} \\
y_{\pi,s0,\myvec{v}} & = & 
\sum \{ \hat{y}_{\pi,s,\myvec{v},\myvec{w}'} \mid \myvec{v} + \myvec{w}' \leq \myvec{C} \}
\label{eq:4} \\
y_{\pi,s1,\myvec{w}} & = & 
\sum \{ \hat{y}_{\pi,s,\myvec{v}',\myvec{w}} \mid \myvec{v}' + \myvec{w} \leq \myvec{C} \}
\label{eq:5} \\
z_{\pi,t0,\myvec{v}} & = & 
\sum \{ \hat{z}_{\pi,t,\myvec{v},\myvec{w}'} \mid \myvec{v} + \myvec{w}' \leq \myvec{C} \}
\label{eq:7} \\
z_{\pi,t1,\myvec{w}} & = & 
\sum \{ \hat{z}_{\pi,t,\myvec{v}',\myvec{w}} \mid \myvec{v}' + \myvec{w} \leq \myvec{C} \}
\label{eq:8} \\
1 & \leq & 
\sum \{ y_{\pi',\epsilon,\myvec{u}'} \mid 
        \mbox{$\pi'$ a 1-type, $\myvec{u}' \leq \myvec{C}$} \}.
\label{eq:1}
\end{eqnarray}

Let $\cE_2$ consist of the following constraints, where $\lambda$,
$\pi$ vary over their standard ranges, $s$, $t$ vary over bit-strings
such that $|s| = p$, $|t| = q$, and $\myvec{u}$ varies over vectors
such that $\myvec{0} < \myvec{u} \leq \myvec{C}$:
\begin{eqnarray}
y_{\pi,s,\myvec{u}} & = & 
\sum \{ x_{\lambda'} \mid \lambda' \in \Lambda_{\pi,s} \mbox{ and } 
        \myvec{C}_{\lambda'} = \myvec{u} \} 
\label{eq:9}\\ 
z_{\pi,t,\myvec{u}} & = & 0 
\qquad \text{\begin{minipage}{6cm}
whenever $\myvec{u}$ is not a scalar multiple
of $\myvec{C}_\tau$ for $\tau = \mu_{\pi,t}$
\end{minipage}}
\label{eq:10}\\
x_{(\lambda^{-1})} & = & x_\lambda 
\label{eq:11}\\
x_\lambda & = & 0
 \qquad \text{whenever ${\rm tp}_1(\lambda) = {\rm tp}_2(\lambda)$}
\label{eq:12}\\
x_\lambda & = & 0 
\qquad \text{whenever $\lambda$ is forbidden}
\label{eq:14} \\
z_{\pi,t,\myvec{u}} & = & 0
 \qquad \text{whenever $\mu_{\pi,t}$ is forbidden.}
\label{eq:15}    
\end{eqnarray}
\noindent
Note that, in~\eqref{eq:10}, $\tau = \mu_{\pi,t}$ is a 2-type, and the
vector $\myvec{C}_{\tau}$ is defined according to~\eqref{eq:Ctau}.  If
the integer encoded by $t$ is less than $R$, $\tau = \mu_{\pi,t}$ will
be a (non-invertible) message-type, and we will have $\myvec{C}_\tau >
\myvec{0}$.  If, on the other hand, the integer encoded by $t$ is
greater than or equal to $R$, $\tau$ will be a silent 2-type, and we
will have $\myvec{C}_\tau = \myvec{0}$.  In this latter case, no
vector $\myvec{u}$ such that $\myvec{u} > \myvec{0}$ can be a multiple
of $\myvec{C}_\tau$, whence, $\cE_2$ contains the constraint
$z_{\pi,t,\myvec{u}} = 0$ for all $\myvec{u}$ such that $\myvec{0} <
\myvec{u} \leq \myvec{C}$ .

Let $\cE_3$ consist of the following constraints, where $\pi$ varies
over all 1-types, $t$ varies over bit-strings such that $|t| = q$, and
$\myvec{u}$ varies over vectors such that $\myvec{0} < \myvec{u} \leq
\myvec{C}$:
\begin{eqnarray}
z_{\pi,t,\myvec{u}} > 0 & \Rightarrow &
\sum\{y_{\pi',\epsilon,\myvec{u}'} \mid 
\mbox{$\pi'= {\rm tp}_2(\mu_{\pi,t})$ and 
$\myvec{u}' \leq \myvec{C}$}\}
 \geq 3mC.  
\label{eq:13}
\end{eqnarray}
\noindent
Again, if the integer encoded by $t$ is greater than or equal to $R$,
we have already argued that $\cE_2$ contains the constraint,
$z_{\pi,t,\myvec{u}} = 0$ for all $\myvec{u}$ such that $\myvec{0} <
\myvec{u} \leq \myvec{C}$, rendering the corresponding instances
of~\eqref{eq:13} trivial.

\vspace{0.25cm}

Finally, let $\cE = \cE_1 \cup \cE_2 \cup \cE_3$.

\section{Main result}
\label{sec:main}
\begin{lemma}
Let $\phi$ and $\cE$ be as above. If $\phi$ is finitely satisfiable,
then $\cE$ has a solution over $\N$.
\label{lma:necessary}
\end{lemma}
\begin{proof}
Suppose $\phi$ is finitely satisfiable. By Lemma~\ref{lma:chromatic},
let $\fA'$ be a finite, chromatic model of $\phi$, and by
Lemma~\ref{lma:dup}, let $\fA = 3mC \cdot \fA'$. Thus, $\fA$ is also
chromatic.  Let $\fA$ have domain $A$. If $\pi$ is a 1-type, let
$A_\pi = \{ a \in A \mid {\rm tp}^\fA[a] = \pi \}$. Now suppose $a \in
A_\pi$. For any bit-string $s$ ($0 \leq |s| \leq p$), define the
$s$-{\em spectrum} of $a$, denoted ${\rm sp}_s^\fA[a]$, to be the
vector whose $i$th component ($1 \leq i \leq m$) is given by
\begin{equation*}
| \{ b \in A : \mbox{$\fA \models f_i[a,b]$, $b \neq a$ \mbox{ and }
     ${\rm tp}^\fA[a,b] \in \Lambda_{\pi,s}$} \} |.
\end{equation*}
For any bit-string $t$ ($0 \leq |t| \leq q$), define the
$t$-{\em tally} of $a$, denoted ${\rm tl}_t^\fA[a]$, to be the vector
whose $i$th component ($1 \leq i \leq m$) is given by
\begin{equation*}
| \{ b \in A : \mbox{$\fA \models f_i[a,b]$, $b \neq a$  \mbox{ and } 
     ${\rm tp}^\fA[a,b] \in M_{\pi,t}$} \} |.
\end{equation*}
These quantities are easiest to understand when the strings $s$ and
$t$ take the value $\epsilon$. For any $a \in A$, ${\rm
sp}_\epsilon^\fA[a]$ is the vector whose $i$th component records the
number of elements $b$ to which $a$ sends a message of {\em
invertible} type containing the atom $f_i(x,y)$. Likewise, ${\rm
tl}_\epsilon^\fA[a]$ is the vector whose $i$th component records the
number of elements $b$ to which $a$ sends a message of {\em
non-invertible} type containing the atom $f_i(x,y)$. If ${\rm
tp}^\fA[a] = \pi$ and $0 < |s| \leq p$, then ${\rm sp}_s^\fA[a]$ is
obtained in the same way as ${\rm sp}_\epsilon^\fA[a]$, except that we
discount all messages whose type is not a member of
$\Lambda_{\pi,s}$. Likewise, if ${\rm tp}^\fA[a] = \pi$ and $0 < |t|
\leq q$, then ${\rm tl}_t^\fA[a]$ is obtained in the same way as ${\rm
tl}_\epsilon^\fA[a]$, except that we discount all messages whose type
is not a member of $M_{\pi,t}$. It is then easy to see that, for all
$a \in A$, all bit-strings $s$ ($|s| < p$) and all bit-strings $t$
($|t| < q$),
\begin{eqnarray}
{\rm sp}_\epsilon^\fA[a] + {\rm tl}_\epsilon^\fA[a] & = & \myvec{C}
\label{nec:1} \\
{\rm sp}_{s0}^\fA[a] + {\rm sp}_{s1}^\fA[a] & = & 
{\rm sp}_s^\fA[a] 
\label{nec:2} \\
{\rm tl}_{t0}^\fA[a] + {\rm tl}_{t1}^\fA[a] & = & 
{\rm tl}_t^\fA[a].
\label{nec:3} 
\end{eqnarray}
Let $\theta: V \rightarrow \N$ be
defined by:
\begin{eqnarray*}
\theta(x_\lambda) & = & |\{ a \in A :
   \mbox{ there exists $b \in A \setminus \{a\}$ such that ${\rm tp}^\fA[a,b] 
          = \lambda$} \} | \\
\theta(y_{\pi,s,\myvec{u}}) & = & |\{ a \in A_\pi :
   {\rm sp}_s^\fA[a] = \myvec{u} \}| \\
\theta(z_{\pi,t,\myvec{u}}) & = & |\{ a \in A_\pi :
   {\rm tl}_t^\fA[a] = \myvec{u} \}| \\
\theta(\hat{y}_{\pi,s,\myvec{v},\myvec{w}}) & = & |\{ a \in A_\pi :
   {\rm sp}_{s0}^\fA[a] = \myvec{v} \mbox{ and }
   {\rm sp}_{s1}^\fA[a] = \myvec{w} \}| \ \ \ \text{whenever $|s| < p$}\\
\theta(\hat{z}_{\pi,t,\myvec{v},\myvec{w}}) & = & |\{ a \in A_\pi : 
   {\rm tl}_{t0}^\fA[a] = \myvec{v} \mbox{ and }
   {\rm tl}_{t1}^\fA[a] = \myvec{w} \}| \ \ \ \ \ \/ \text{whenever $|t| < q$},
\end{eqnarray*}
where the indices $\lambda$, $\pi$, $s$, $t$, $\myvec{u}$, $\myvec{v}$
and $\myvec{w}$ vary over their standard ranges.

We show that $\theta$ is a solution of $\cE$ by checking the
forms~\eqref{eq:2}--\eqref{eq:13} in turn.  The
constraints~\eqref{eq:2}--\eqref{eq:6} follow easily
from Equations~\eqref{nec:1}--\eqref{nec:3},
respectively.  The constraints~\eqref{eq:4}--\eqref{eq:8} are
immediate.  In the (single) constraint~\eqref{eq:1}, the sum on the
right-hand side evaluates under $\theta$ to the cardinality of $A$,
which is positive by definition. To understand the
constraints~\eqref{eq:9}, fix any $\pi$ and $s$ (with $|s| = p$), and
write, for any $\lambda \in \Lambda_{\pi,s}$,
\begin{equation*}
A_{\lambda} =  \{ a \in A \mid 
   \mbox{ there exists $b \in A \setminus \{a\}$ such that ${\rm tp}^\fA[a,b] 
          = \lambda$} \}.
\end{equation*}
Thus, $A_{\lambda} \subseteq A_{\pi}$, and $|A_{\lambda}| =
\theta(x_{\lambda})$.  By Lemma~\ref{lma:onlyone}, for any $a \in
A_\pi$, there can be at most one element $b \in A \setminus
\{a\}$ such that ${\rm tp}^\fA[a,b] \in \Lambda_{\pi,s}$.  It follows
that the sets $A_{\lambda}$, for 
$\lambda$ varying over $\Lambda_{\pi,s}$, are
pairwise disjoint, and, moreover, that $a \in A_\pi$ has $s$-spectrum
$\myvec{u} > \myvec{0}$ if any only if it is a member of some (hence,
exactly one) of these sets $A_{\lambda}$, with $\myvec{C}_{\lambda} =
\myvec{u}$. That is, for all $\myvec{u} > \myvec{0}$, 
\begin{equation*}
\{ a \in A_\pi \mid {\rm sp}_s^\fA[a] =
\myvec{u} \} = \bigcup \{ A_{\lambda'} \mid
\lambda' \in \Lambda_{\pi,s} \mbox{ and } 
\myvec{C}_{\lambda'} =
\myvec{u} \},
\end{equation*}
with the sets on the right-hand side pairwise disjoint.  The relevant
instance of~\eqref{eq:9} is then immediate from the definition of
$\theta$.  To see why the constraints~\eqref{eq:10} hold, note that,
if $|t| = q$, then $M_{\pi,t}$ is a set containing precisely one
2-type $\tau = \mu_{\pi,t}$, which is either a non-invertible
message-type or a silent 2-type. Either way, for every $a \in A_\pi$,
${\rm tl}_{t}^\fA[a]$ must be a scalar multiple (possibly zero) of
$\myvec{C}_\tau$. In other words, if $\myvec{u} > \myvec{0}$ is not
a scalar multiple of $\myvec{C}_\tau$, then
$\theta(z_{\pi,t,\myvec{u}}) = 0$.  To understand the
constraints~\eqref{eq:11}, observe that, since $\fA$ is chromatic,
$\theta(x_\lambda)$ is actually the total number of messages of
(invertible) type $\lambda$ sent by elements of $\fA$, and similarly
for $\theta(x_{(\lambda^{-1})})$; and these numbers are obviously
equal. The constraints~\eqref{eq:12} are immediate given that $\fA$ is
chromatic. The constraints \eqref{eq:14} and~\eqref{eq:15} are
immediate given that $\fA \models \phi$. To understand the
constraints~\eqref{eq:13}, fix any $\pi$, $t$ and $\myvec{u}$ such
that $|t| = q$ and $\myvec{u} > \myvec{0}$.  If $z_{\pi,t,\myvec{u}} >
0$ holds under $\theta$, then $\mu_{\pi,t}$ is a (non-invertible)
message-type; and furthermore, at least one message of that type must
be sent in $\fA$, so that $\fA$ contains at least one element whose
1-type is ${\rm tp}_2(\mu_{\pi,t})$ and hence---since $\fA = 3mC \cdot
\fA'$---at least $3mC$ such elements. But the exact number of 
elements in $A$ whose 1-type is ${\rm tp}_2(\mu_{\pi,t})$
is given by the value, under $\theta$, of
$\sum\{y_{\pi',\epsilon,\myvec{u}'} \mid \mbox{$\pi'= {\rm
tp}_2(\mu_{\pi,t})$ and $\myvec{u}' \leq \myvec{C}$}\}$.
\end{proof}

In establishing the converse of Lemma~\ref{lma:necessary}, the
following technical result concerning solutions of $\cE$ will prove
useful.  To avoid notational clutter, we use the variable names
$x_\lambda$, $y_{\pi,s,\myvec{u}}$, $z_{\pi,t,\myvec{u}}$,
$\hat{y}_{\pi,s,\myvec{v},\myvec{w}}$,
$\hat{z}_{\pi,t,\myvec{v},\myvec{w}}$ to stand for the corresponding
natural numbers in some such solution (and similarly for terms
involving these variables).
\begin{lemma}
Let $x_\lambda$, $y_{\pi,s,\myvec{u}}$, $z_{\pi,t,\myvec{u}}$,
$\hat{y}_{\pi,s,\myvec{v},\myvec{w}}$,
$\hat{z}_{\pi,t,\myvec{v},\myvec{w}}$ $($with indices having the
appropriate ranges$)$ be natural numbers satisfying the constraints
$\cE$ given above.  Fix any 1-type $\pi$, and let $A_\pi$ be a set of
cardinality $\sum \{y_{\pi, \epsilon, \myvec{u'}} \mid \myvec{u'} \leq
\myvec{C} \}$. Then there exists a system of functions on $A_\pi$
\[
f_{\pi,s}: A_\pi \rightarrow \{ \myvec{u} \mid \myvec{u} \leq \myvec{C} \}
\hspace{1cm}
g_{\pi,t}: A_\pi \rightarrow \{ \myvec{u} \mid \myvec{u} \leq \myvec{C} \},
\]
where the indices $s$ and $t$ vary over their standard ranges, such
that, for all vectors $\myvec{u} \leq \myvec{C}$,
\begin{eqnarray}
| f_{\pi,s}^{-1}(\myvec{u}) | & = & y_{\pi,s,\myvec{u}}
\label{suf:1f}\\
| g_{\pi,t}^{-1}(\myvec{u}) | & = & z_{\pi,t,\myvec{u}},
\label{suf:1g}
\end{eqnarray}
and such that, for all $a \in A_\pi$,
\begin{eqnarray}
\sum \{f_{\pi,s'}(a) : |s'| = p \} + 
\sum \{g_{\pi,t'}(a) : |t'| = q \} & = & \myvec{C}.
\label{suf:2}
\end{eqnarray}
\label{lma:fg}
\end{lemma}
\begin{proof}
Decompose the set $A_\pi$ into pairwise disjoint (possibly empty) sets
$A_{\myvec{u}}$ such that $|A_{\myvec{u}}| =
y_{\pi,\epsilon,\myvec{u}}$, where the index $\myvec{u}$ varies over all
vectors $\leq \myvec{C}$. This is possible by the cardinality of
$A_\pi$. For all $\myvec{u} \leq \myvec{C}$, and all $a \in
A_{\myvec{u}}$, set
\begin{equation*}
f_{\pi,\epsilon}(a) = \myvec{u} \hspace{1cm}
g_{\pi,\epsilon}(a) = \myvec{C} - \myvec{u}.
\end{equation*}
This assignment evidently satisfies~\eqref{suf:1f} for $s =
\epsilon$; and by the constraints~\eqref{eq:2}, it also
satisfies~\eqref{suf:1g} for $t = \epsilon$. We observe in passing
that, for all $a \in A_\pi$,
\begin{equation}
f_{\pi,\epsilon}(a) + g_{\pi,\epsilon}(a) =  \myvec{C}.
\label{suff:3}
\end{equation}
We now construct the functions $f_{\pi,s}$, where $0 < |s| \leq p $,
by induction on $s$. Assume that, for some $s$ ($0 \leq |s| < p$),
$f_{\pi,s}$ has been defined and satisfies~\eqref{suf:1f}.  For every
vector $\myvec{u} \leq \myvec{C}$, decompose
$f_{\pi,s}^{-1}(\myvec{u})$ into pairwise disjoint (possibly empty)
sets $A_{\myvec{v}, \myvec{w}}$ such that $|A_{\myvec{v},
  \myvec{w}}| = \hat{y}_{\pi,s,\myvec{v}, \myvec{w}}$, where the indices
$\myvec{v}$, $\myvec{w}$ vary over all vectors satisfying $\myvec{v} +
\myvec{w} = \myvec{u}$. This is possible by the
constraints~\eqref{eq:3} together with the assumption that $f_{\pi,s}$
satisfies~\eqref{suf:1f}.  Having thus decomposed the sets
$f_{\pi,s}^{-1}(\myvec{u})$ (for all $\myvec{u} \leq \myvec{C}$), we
see that, for any $a \in A_\pi$, there is precisely one (ordered) pair
of vectors $\myvec{v}$, $\myvec{w}$ such that 
$a \in A_{\myvec{v},
  \myvec{w}}$; 
hence we may set
\begin{equation*}
f_{\pi,s0}(a) = \myvec{v}  \hspace{1cm}  f_{\pi,s1}(a) = \myvec{w}.
\end{equation*}
This defines the functions $f_{\pi,s0}$ and $f_{\pi,s1}$. 
We observe in passing that, for all $a \in A_\pi$, 
\begin{equation}
f_{\pi,s0}(a) + f_{\pi,s1}(a) =  f_{\pi,s}(a).
\label{suff:4}
\end{equation}
To see that $f_{\pi,s0}$ and $f_{\pi,s1}$ both satisfy
Equation~\eqref{suf:1f}, note that $f_{\pi,s0}(a) = \myvec{v}$ if and
only if, for some vector $\myvec{w}'$ such that $\myvec{v} +
\myvec{w}' \leq \myvec{C}$, $a \in A_{\myvec{v},
\myvec{w}'}$. Similarly, $f_{\pi,s1}(a) = \myvec{w}$ if and only if,
for some vector $\myvec{v}'$ such that $\myvec{v}' + \myvec{w} \leq
\myvec{C}$, $a \in A_{\myvec{v}', \myvec{w}}$. That is,
\begin{eqnarray*}
f_{\pi,s0}^{-1}(\myvec{v}) & = & 
  \bigcup \{ A_{\myvec{v},\myvec{w}'} \mid 
     \myvec{v} + \myvec{w}' \leq \myvec{C} \} \\
f_{\pi,s1}^{-1}(\myvec{w}) & = & 
  \bigcup \{ A_{\myvec{v}',\myvec{w}} \mid 
     \myvec{v}' + \myvec{w} \leq \myvec{C} \},
\end{eqnarray*}
with the collections of sets on the respective right-hand sides being
pairwise disjoint.  By the constraints~\eqref{eq:4}--\eqref{eq:5},
together with the fact that $|A_{\myvec{v},\myvec{w}}| =
\hat{y}_{\pi,s,\myvec{v},\myvec{w}}$ for all $\myvec{v},\myvec{w}$, we
have:
\begin{eqnarray*}
| f_{\pi,s0}^{-1}(\myvec{v}) | & = & y_{\pi,s0,\myvec{v}}\\
| f_{\pi,s1}^{-1}(\myvec{w}) | & = & y_{\pi,s1,\myvec{w}},
\end{eqnarray*}
which establishes \eqref{suf:1f} for the functions $f_{\pi,s0}$ and
$f_{\pi,s1}$. This completes the induction.  The construction of the
functions $g_{\pi,t}$ proceeds completely analogously, using
Constraints~\eqref{eq:6}, \eqref{eq:7} and~\eqref{eq:8}. In carrying
out this latter construction, we obtain, in a parallel way
to~\eqref{suff:4},
\begin{equation}
g_{\pi,t0}(a) + g_{\pi,t1}(a) =  g_{\pi,t}(a),
\label{suff:5}
\end{equation}
for all bit-strings $t$ ($|t| < q$) and all $a \in A_\pi$.

It remains to establish~\eqref{suf:2}. We prove the stronger result
that, for all $a \in A_\pi$, $j$ ($0 \leq j \leq p$) and $k$ ($0 \leq
k \leq q$),
\begin{eqnarray}
\sum \{f_{\pi,s'}(a) : |s'| = j \} + 
\sum \{g_{\pi,t'}(a) : |t'| = k \} & = & \myvec{C},
\label{suf:2i}
\end{eqnarray}
using a double induction on $j$ and $k$.  If $j = k = 0$, then the
left-hand side of~\eqref{suf:2i} is simply $f_{\pi,\epsilon}(a) +
g_{\pi,\epsilon}(a)$, which is equal to $\myvec{C}$ by~\eqref{suff:3}.
Suppose now that the result holds for the pair $j$, $k$, with $j < p$.
Then
\begin{eqnarray*}
\ & \ & \sum \{f_{\pi,s'}(a) : |s'| = (j+1) \} + 
        \sum \{g_{\pi,t'}(a) : |t'| = k \} \\
\ & = & \sum \{f_{\pi,s'0}(a) + f_{\pi,s'1}(a)  : |s'| = j \} + 
        \sum \{g_{\pi,t'}(a) : |t'| = k \} \\
\ & = & \sum \{f_{\pi,s}(a) : |s'| = j \} + 
        \sum \{g_{\pi,t'}(a) : |t'| = k \} \qquad \text{ by~\eqref{suff:4} } \\
\ & = & \myvec{C} \qquad \text{ by inductive hypothesis.} 
\end{eqnarray*}
This establishes the result for the pair $j+1,k$.  An analogous
argument using~\eqref{suff:5} applies when $k < m$, completing the
induction.
\end{proof}
Before we come to the promised converse of Lemma~\ref{lma:necessary},
we remark on the (exponentially many) choices made during the
construction of the various functions $f_{\pi,s}$ and $g_{\pi,t}$ in
the proof of Lemma~\ref{lma:fg}---specifically, in the decomposition
of certain sets into collections of subsets.  It is because of this
large number of independent choices that solutions of $\cE$ typically
encode not one, but many, models of $\phi$.
\begin{lemma}
Let $\phi$ and $\cE$ be as above. If $\cE$ has a solution over $\N$,
then $\phi$ is finitely satisfiable.
\label{lma:sufficient}
\end{lemma}
\begin{proof}
Suppose $\cE$ has a solution over $\N$.  Again, we use the variable
names $x_\lambda$, $y_{\pi,s,\myvec{u}}$, $z_{\pi,t,\myvec{u}}$,
$\hat{y}_{\pi,s,\myvec{v},\myvec{w}}$,
$\hat{z}_{\pi,t,\myvec{v},\myvec{w}}$ to stand for the corresponding
values in some such solution.  Our task is to construct a model $\fA$
of $\phi$.

For each 1-type $\pi$, let $A_\pi$ be a set of cardinality $\sum
\{y_{\pi, \epsilon, \myvec{u}} \mid \myvec{u} \leq \myvec{C} \}$, with
the $A_\pi$ pairwise disjoint; and let $A = \bigcup \{A_\pi \mid
\mbox{$\pi$ a 1-type} \}$.  Think of $A_\pi$ as the set of elements of
$A$ which `want' to have 1-type $\pi$.  By the
constraint~\eqref{eq:1}, $A \neq \emptyset$. For every 1-type $\pi$,
let the functions $f_{\pi,s}$ and $g_{\pi,t}$ be constructed as in
Lemma~\ref{lma:fg}; we are interested only in those $f_{\pi,s}$ and
$g_{\pi,t}$ where $|s| = p$, and $|t| = q$. For all such $\pi$, $s$,
$t$, and all $a \in A_\pi$, think of $f_{\pi,s}(a)$ as the
$s$-spectrum which $a$ `wants' to have, and think of $g_{\pi,t}(a)$ as
the $t$-tally which $a$ `wants' to have.  Finally, consider any set
$f_{\pi,s}^{-1}(\myvec{u})$, where $\myvec{0} < \myvec{u} \leq
\myvec{C}$ and $|s| = p$.  Using the constraints~\eqref{eq:9} and
Equation~\eqref{suf:1f}, we can decompose $f_{\pi,s}^{-1}(\myvec{u})$
into pairwise disjoint (possibly empty) sets $A_{\lambda}$ with
$|A_{\lambda}| = x_{\lambda}$, where $\lambda$ varies over the set of
invertible message-types such that $\lambda \in \Lambda_{\pi,s}$ and
$\myvec{C}_{\lambda} = \myvec{u}$.  It follows that, if $a \in
A_\lambda$, with $\lambda \in \Lambda_{\pi,s}$, then $\myvec{C}_\lambda
= f_{\pi,s}(a)$.  Think of $A_{\lambda}$ as the set of elements of
$A_\pi$ which `want' to send a single message of (invertible) type
$\lambda$.

Before proceeding, we pause to consider the construction just
described in respect of any of the sets $A_\pi$. Fixing, for the
moment, some bit-string $s$ with $|s| = p$, we see that $A_\pi$ is
decomposed into the pairwise disjoint sets $f_{\pi,s}^{-1}(\myvec{u})$
(as $\myvec{u}$ varies over vectors such that $\myvec{u} \leq
\myvec{C}$), and that each of the sets $f_{\pi,s}^{-1}(\myvec{u})$,
where $\myvec{0} < \myvec{u} \leq \myvec{C}$, is further decomposed
into the pairwise disjoint subsets $A_\lambda$ (as $\lambda$ varies
over the elements of $\Lambda_{\pi,s}$ such that $\myvec{C}_\lambda =
\myvec{u}$). Note that the set $f_{\pi,s}^{-1}(\myvec{0})$ is not
subject to this further stage of decomposition.  This process is
performed for {\em every} bit string $s$ with $|s| = p$, so that
different values of $s$ lead to independent---and possibly
overlapping---decompositions, as illustrated in
Fig.~\ref{fig:decomposition}. Likewise, for every bit-string $t$ with
$|t| = q$, $A_\pi$ is decomposed into the pairwise disjoint sets
$g_{\pi,t}^{-1}(\myvec{u})$ (as $\myvec{u}$ varies over vectors such
that $\myvec{u} \leq \myvec{C}$). Again, decompositions corresponding
to different values of $t$ should be thought of as independent of each
other.
\begin{figure}[ht]
\begin{flushleft}
\hspace{0.2cm}
\input{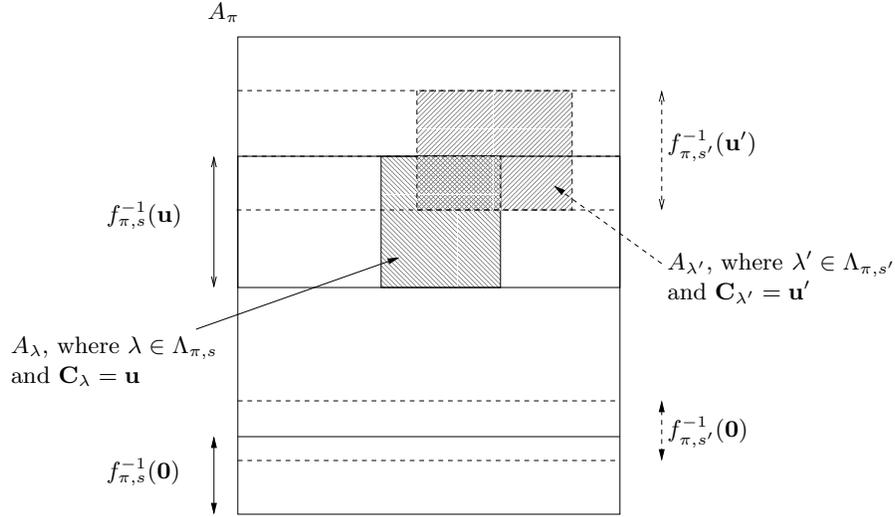}
\end{flushleft}
\caption{The decompositions of $A_\pi$  for the strings $s$ and $s'$.}
\label{fig:decomposition}
\end{figure}

We now proceed to construct, for every $a \in A$, a data-structure
recording a collection of messages sent by $a$, each of which is
labelled with some (invertible or non-invertible) message-type.
(These data-structures will later be combined to form a structure
$\fA$ on $A$.)  Recall that, if $\pi$ is any 1-type, then
$\mu_{\pi,0}, \ldots, \mu_{\pi,R-1}$ is an enumeration of the
non-invertible message-types $\mu$ such that ${\rm tp}_1(\mu) = \pi$.
Fix $a \in A$, and let $\pi$ be the unique 1-type such that $a \in
A_\pi$. The messages sent by $a$ shall be as follows.  (i) For every
bit-string $s$ such that $|s| = p$, if $f_{\pi,s}(a) > \myvec{0}$, let
$\lambda_{a,s}$ be the invertible message-type $\lambda \in
\Lambda_{\pi,s}$ such that $a \in A_{\lambda}$ (hence
$\myvec{C}_{\lambda} = f_{\pi,s}(a)$), and let $a$ send a single
message labelled $\lambda_{a,s}$.  Note that, if $f_{\pi,s}(a) >
\myvec{0}$, then $\lambda_{a,s}$ exists and is unique by the
construction of the sets $A_\lambda$. (ii) For every bit string $t$
such that $|t| = q$, if $t$ encodes an integer less than $R$ (so that
$\mu = \mu_{\pi,t}$ is a non-invertible message-type and
$\myvec{C}_\mu > \myvec{0}$), let $n_{a,t}$ be the unique natural
number $n$ such that $g_{\pi,t}(a) = n\myvec{C}_\mu$, and let $a$ send
$n_{a,t}$ distinct messages labelled $\mu_{\pi,t}$.  Note that, if
$g_{\pi,t}(a) = \myvec{0}$, then $n_{a,t} = 0$; on the other hand, if
$g_{\pi,t}(a) > \myvec{0}$, then $n_{a,t}$ exists by the
constraints~\eqref{eq:10} and Equation~\eqref{suf:1g}.  The resulting
data-structure is depicted in Fig.~\ref{fig:messages}, where, for
readability, we have replaced any bit-strings by the integers they
conventionally denote.
\begin{figure}[ht]
\centerline{\input{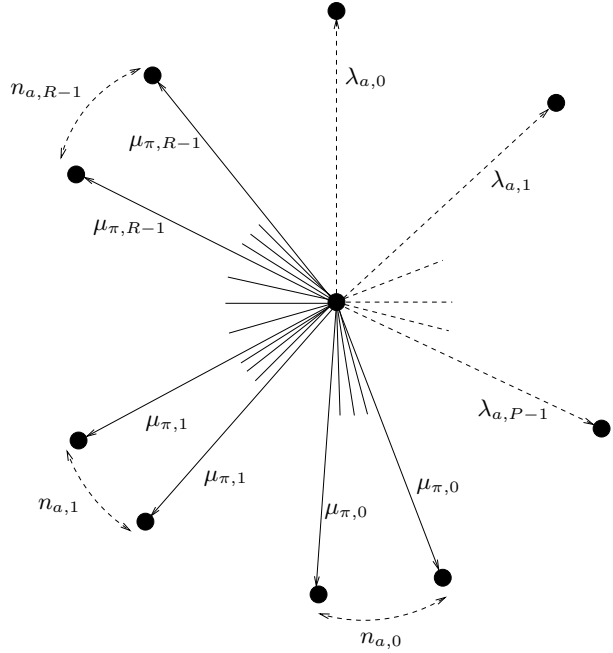}}
\caption{The messages sent by $a \in A_\pi$. For each $j$ ($0 \leq j <
P$), $a$ may or may not send a message labelled $\lambda_{a,j}$ (hence
the dotted lines); if it does, then $\lambda_{a,j} \in
\Lambda_{\pi,j}$. For each $k$ ($0 \leq k < R$), $a$ sends $n_{a,k}$
messages labelled $\mu_{\pi,k}$; but the numbers $n_{a,k}$ can be zero.}
\label{fig:messages}
\end{figure}
For all $a \in A$ and all $i$ ($1 \leq i \leq m$), let $C_{a,i}$ ($1
\leq i \leq m$) be the number of messages sent by $a$ having any label
$\nu$ for which $f_i(x,y) \in \nu$, and furthermore let $\myvec{C}_a$
be the vector $(C_{a,1}, \ldots, C_{a,m})$. By inspection of
Fig.~\ref{fig:messages}, 
\begin{equation*}
\myvec{C}_a =
\sum \{f_{\pi,s'}(a) : |s'| = p \} + 
\sum \{g_{\pi,t'}(a) : |t'| = q \},
\end{equation*}
and so, by Equation~\eqref{suf:2},
\begin{equation}
\myvec{C}_a =  \myvec{C}.
\label{eq:count}
\end{equation}

We now build $\fA$ in four steps as follows.

\vspace{0.25cm}

\noindent
{\bf Step 1} (Fixing the 1-types): For all 1-types $\pi$ and all $a
\in A_\pi$, set ${\rm tp}^\fA[a] = \pi$. Since the $A_\pi$ are
pairwise disjoint, no clashes arise.

\vspace{0.25cm}

\noindent
{\bf Step 2} (Fixing the invertible message-types): Let $\lambda$ be
any invertible message-type. By construction, exactly $|A_\lambda | =
x_\lambda$ elements of $A$ send some message labelled with $\lambda$,
and each of those elements sends exactly one such message.  Hence, the
number of messages labelled with $\lambda$ (over all $a \in A$) is
$x_\lambda$; likewise, the number of messages labelled with
$\lambda^{-1}$ is $x_{\lambda^{-1}}$. By the
constraints~\eqref{eq:11}, we may put the $\lambda$-labelled messages
and the $\lambda^{-1}$-labelled messages in 1--1 correspondence. If $a
\in A$ sends a $\lambda$-labelled message, let $b \in A$ send the
corresponding $\lambda^{-1}$-labelled message, and set ${\rm
tp}^\fA[a,b] = \lambda$.  For this assignment to make sense, we need
to check that $a$ and $b$ are distinct. But, by construction, we must
have $x_\lambda > 0$, whence, by the constraints~\eqref{eq:12}, ${\rm
tp}_1(\lambda) \neq {\rm tp}_2(\lambda)$, so that $A_{{\rm
tp}_1(\lambda)}$ and $A_{{\rm tp}_2(\lambda)}$ are disjoint sets
containing $a$ and $b$, respectively.  Thus, the assignment ${\rm
tp}^\fA[a,b] = \lambda$ makes sense, and does not clash the with
1-type assignments in Step~1.  We can think of the element $b$ as
`receiving' the message sent by $a$ (and vice versa).  Moreover, by
construction, for every 1-type $\pi'$, $a$ sends at most one message
labelled with an invertible message-type $\lambda'$ such that ${\rm
tp}_2(\lambda') = \pi'$. Therefore, there is no chance that these
assignments clash with each other.

\vspace{0.25cm}

\noindent
{\bf Step 3} (Fixing the non-invertible message-types): As a
preliminary, for every 1-type $\pi$, we decompose $A_\pi$ into three
pairwise disjoint (possibly empty) sets $A_{\pi,0}$, $A_{\pi,1}$ and
$A_{\pi,2}$ satisfying the condition that, if $|A_\pi| \geq 3mC$, then
$|A_{\pi,j}| \geq mC$ for all $j$ ($0 \leq j \leq 2$). Now let $\mu$
be any non-invertible message-type, let $\pi = {\rm tp}_1(\mu)$, and
let $\rho = {\rm tp}_2(\mu)$. (Note that $\pi$ and $\rho$ may be
identical.) Let $t$ be the bit-string of length $q$ such that $\mu =
\mu_{\pi,t}$, and suppose some element $a$ sends $n_{a,t} > 0$
messages labelled $\mu$.  It follows that $a \in A_\pi$, and also that
there is a vector $\myvec{u} > \myvec{0}$ such that $g_{\pi,t}(a) =
\myvec{u}$, and hence such that $g_{\pi,t}^{-1}(\myvec{u})$ is
non-empty.  By Equation~\eqref{suf:1g}, $z_{\pi,t,\myvec{u}} >0$,
whence, by the constraints~\eqref{eq:13}, $\sum \{
y_{\rho,\epsilon,\myvec{u}'} \mid \myvec{u}' \leq \myvec{C} \} \geq
3mC$. But recall that, since $\rho$ is a 1-type, $|A_\rho| = \sum \{
y_{\rho,\epsilon,\myvec{u}'} \mid \myvec{u}' \leq \myvec{C} \}$, so
that each of the sets $A_{\rho, 0}$, $A_{\rho, 1}$ and $A_{\rho, 2}$
contains at least $mC$ elements. Since $a \in A_\pi$, let $j$ ($0 \leq
j \leq 2$) be such that $a \in A_{\pi,j}$, let $k = j+1$ (mod 3), and
select $n_{a,t}$ elements $b$ from $A_{\rho,k}$ which have not yet
been chosen to receive any other messages (invertible or
non-invertible) sent by $a$. Since the total number of messages sent
by $a$ is certainly at most $mC$, we never run out of choices. For
each of these elements $b$, set ${\rm tp}^\fA[a,b] = \mu$. Since $\pi
= {\rm tp}_1(\mu)$ and $\rho = {\rm tp}_2(\mu)$, these assignments
cannot clash with those made in Step~1, and by construction, they
cannot clash with assignments corresponding to other messages sent by
$a$.  We need only check that they cannot clash with assignments
corresponding to messages sent by $b$. Specifically, we must ensure
that, if ${\rm tp}^\fA[a,b] = \mu$ is assigned as just described, it
is not possible for $a$ to be chosen to receive a $\mu'$-labelled
message sent by $b$, where $\mu'$ is some non-invertible message-type.
But any $\mu'$-labelled message sent by $b \in A_{\rho,k}$, with ${\rm
tp}_2(\mu') = \pi$, could only be sent to an element in $A_{\pi,j'}$,
where $j' = k+1$ (mod 3); and by assumption, $A_{\pi,j}$ and
$A_{\pi,j'}$ are disjoint, (Fig.~\ref{fig:noninvertible}). Observe
that this conclusion follows even if $\pi = \rho$.
\begin{figure}
\begin{center}
\begin{picture}(100,80)(0,0)

\put(10,10){\framebox(40,60)}
\put(10,30){\dashbox{2}(40,0){}}
\put(10,50){\dashbox{2}(40,0){}}

\put(90,10){\framebox(40,60)}
\put(90,30){\dashbox{2}(40,0){}}
\put(90,50){\dashbox{2}(40,0){}}

\put(30,60){\vector(4,-1){79}}
\put(110,40){\vector(-4,-1){79}}

\put(23,58){$a$}
\put(113,38){$b$}

\put(30,60){\circle*{1.5}}
\put(30,20){\circle*{1.5}}
\put(110,40){\circle*{1.5}}

\put(28,0){$A_\pi$}
\put(108,0){$A_\rho$}
\put(-14,56){$A_{\pi,j}$}
\put(-14,16){$A_{\pi,j'}$}
\put(136,36){$A_{\rho,k}$}
\end{picture}
\end{center}
\caption{Fixing the non-invertible message-types.}
\label{fig:noninvertible}
\end{figure}
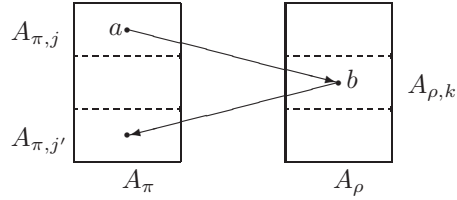

\vspace{0.25cm}

\noindent
{\bf Step 4} (Fixing the remaining 2-types): Recall that a guard-atom
is any atom $p(x,y)$ or $p(y,x)$, where $p$ is a binary predicate.  
If ${\rm tp}^\fA[a,b]$ has not been defined, set it to be
the 2-type
\[
\pi \cup \rho[y/x] \cup
\{ \neg \gamma \mid \text{ $\gamma$ is a guard-atom not involving $\approx$} \},
\]
where $\pi = {\rm tp}^\fA[a]$, $\rho = {\rm tp}^\fA[b]$, and
$\rho[y/x]$ is the result of replacing $x$ by $y$ in $\rho$.
Note that
neither this 2-type nor its inverse is a message-type. Note also that,
since $a$ and $b$ certainly send some messages, the
constraints~\eqref{eq:14} and~\eqref{eq:15} ensure that both $\alpha \wedge
\bigwedge \pi$ and $\alpha \wedge \bigwedge \rho$ are satisfiable.

This completes the definition of $\fA$; it remains to show that $\fA
\models \phi$. Referring to~\eqref{eq:normalform}, we consider first
the conjuncts:
\begin{equation*}
\forall x \alpha \wedge 
 \bigwedge_{1 \leq h \leq l} \forall x \forall y  
     (e_h(x,y) \rightarrow (\beta_h \vee x \approx y)).
\end{equation*}
We see from the constraints~\eqref{eq:14} and~\eqref{eq:15} that no 2-type
assignment in Steps~2 and~3 violates these conjuncts. And it is obvious
that no assignment in Step~4 does so. (This is where we use the
guardedness of $\phi$, of course.) Finally, we consider the conjuncts
\begin{equation*}
   \bigwedge_{1 \leq i \leq m} 
      \forall x \exists_{=C_i} y (f_i(x,y) \wedge x \not \approx y).
\end{equation*}
To see that these conjuncts are all satisfied, it suffices
to note Equation~\eqref{eq:count} and the
fact that none of the 2-types assigned in Step~4 is a message-type.
We remark in passing that $\fA$ is chromatic.
\end{proof}

The constraints $\cE$ all have the forms
\begin{equation}
\begin{array}{rcl}
x_1 + \cdots + x_n  & = & x\\
x_1 + \cdots + x_n  & \geq & 1\\
x & = & 0\\
x > 0 \ \  \Rightarrow \ \  x_1 + \cdots + x_n & \geq & D,
\end{array}
\label{eq:conds}
\end{equation}
where $n >0$, $x, x_1, \ldots, x_n$ are variables, and $D$ is a positive
integer.  We measure the size $\lVert \cE \rVert$ of $\cE$ in the
usual way, with binary encoding of the constants $D$.

The following lemma essentially repeats Lutz, 
Sattler and Tendera~\cite{logic:lst05}, Proposition~11. (Those
authors in turn credit Calvanese~\cite{logic:calvanesePhD}.) We
repeat the proof for convenience.
\begin{lemma}
Let $\phi$ and $\cE$ be as above.  An algorithm exists to determine,
in time bounded by an exponential function of $\lVert \phi \rVert$,
whether $\cE$ has a solution over $\N$.
\label{lma:exp}
\end{lemma}
\begin{proof}
Suppose $\cE$ has a solution $\theta: V \rightarrow \N$. We define the
integer programming problem $\cE_{\theta}$ by replacing every
constraint in $\cE$ having the form $x > 0 \Rightarrow x_1 + \cdots +
x_n \geq D$ with one of two corresponding inequalities as follows:
\begin{equation*}
\begin{cases}
x = 0 \qquad \qquad \qquad \qquad \text{ if $\theta(x) = 0$}\\
x_1 + \cdots + x_n \geq D \qquad \text{ \ otherwise.}
\end{cases}
\end{equation*}
It is easy to check that $\theta$ is a solution of $\cE_{\theta}$,
and, moreover, that every solution of $\cE_{\theta}$ is a solution of
$\cE$. Since $\cE_{\theta}$ is an integer programming problem and has
a solution over $\N$, by a well-known theorem
(Papadimitriou~\cite{logic:papadimitriou81}), it has a solution
$\theta'$ over $\N$ in which every value is bounded by a (positive)
integer $H$, where $H$ can be computed (as a binary string) in time
bounded by a polynomial function of $\lVert \cE_{\theta} \rVert$ and
hence in time bounded by an exponential function of $\lVert \phi
\rVert$.  (Of course, the {\em integer} $H$ is bounded only by an {\em
doubly} exponential function of $\lVert \phi \rVert$.) Moreover,
$\theta'$ must also be a solution of $\cE$.  Therefore $\cE$ too has a
solution over $\N$ if and only if it has a solution over $\N$
in which every value is
bounded by $H$.

Now consider the integer programming problem $\cE_H$ defined by
replacing every constraint of the form $x > 0 \Rightarrow x_1 + \cdots
+ x_n \geq D$ in $\cE$ by the corresponding inequalities
\begin{eqnarray*}
Hy & \geq & x \\
x_1 + \cdots + x_n & \geq & Dy,
\end{eqnarray*}
where $y$ is a new variable. Every solution of $\cE_H$ over $\N$ is a
solution of $\cE$. Moreover, suppose $\theta'$ is any solution of
$\cE$ over $\N$ in which all values are bounded by $H$; and let $y$ be
one of the new variables of $\cE_H$, introduced to eliminate the
constraint $x > 0 \Rightarrow x_1 + \cdots + x_n \geq D$. Let us
extend $\theta'$ to give a value to $y$ as follows:
\begin{equation*}
\theta'(y) = 
\begin{cases}
0 \qquad \text{ if $\theta'(x) = 0$}\\
1 \qquad \text{ otherwise.}
\end{cases}
\end{equation*}
It is routine to check that extending $\theta'$ in this way for all
the new variables $y$ in $\cE_H$ yields a solution of $\cE_H$.  Hence
$\cE$ can be transformed, in time bounded by an exponential function
of $\lVert \phi \rVert$, into the equisatisfiable (over $\N$)
constraint set $\cE_H$, in which all constraints are of the forms
\begin{equation*}
\begin{array}{rclcrcl}
x_1 + \cdots + x_n  & = & x & \hspace{1cm} & 
x & = & 0\\
x_1 + \cdots + x_n  & \geq & 1
& \hspace{1cm} & 
Dx_1  & \geq & x_2\\
x_1 + \cdots + x_n  & \geq & Dx,
\end{array}
\end{equation*}
where, again, the $D$ are positive integers.
It is obvious that, if $\cE_H$ has a solution over the non-negative
rationals, then it has a solution over $\N$ as well. (Simply multiply
by the product of all the denominators.) Hence, we can
equivalently regard $\cE_H$ as a linear programming problem. But
linear programming is in PTIME, by Khachiyan's
theorem~\cite{logic:khachiyan79}.
\end{proof}
\begin{theorem}
The finite satisfiability problem for $\cGC^2$ is in EXPTIME.
\label{theo:main}
\end{theorem}
\begin{proof}
Lemmas~\ref{lma:normalform}, \ref{lma:necessary}, \ref{lma:sufficient}
and~\ref{lma:exp}.
\end{proof}
\section{The Satisfiability Problem}
\label{sec:dessert}
The above technique also provides a simple proof of a result derived in
Kazakov~\cite{logic:kazakov04}, namely, that the satisfiability
problem for $\cGC^2$ is in EXPTIME.
\begin{notation}
  Let $\N^*$ denote the set $\N \cup \{ \aleph_0 \}$.  We extend the
  ordering $>$ and the arithmetic operations $+$ and $\cdot$ from $\N$
  to $\N^*$ in the obvious way. Specifically, we define $\aleph_0 > n$
  for all $n \in \N$; we define $\aleph_0 + \aleph_0 = \aleph_0 \cdot
  \aleph_0 = \aleph_0$ and $0 \cdot \aleph_0 = \aleph_0 \cdot 0 = 0$;
  we define $n + \aleph_0 = \aleph_0 +n = \aleph_0$ for all $n \in
  \N$; and we define $n \cdot \aleph_0 = \aleph_0 \cdot n = \aleph_0$
  for all $n \in \N$ such that $n >0$.  Under this extension, $>$
  remains a total order, and $+$, $\cdot$ remain associative and
  commutative.
\label{notation:arithmetic}
\end{notation}
Consider again the constraints $\cE$ given
in~\eqref{eq:2}--\eqref{eq:13}, but now with the variables ranging
over the whole of $\N^*$. Using the arithmetic in
Notation~\ref{notation:arithmetic}, the reasoning of
Lemmas~\ref{lma:necessary}--\ref{lma:sufficient} works
unproblematically even when countably infinite sets are allowed. Thus,
we have:
\begin{lemma}
Let $\phi$ and $\cE$ be as above. Then $\phi$ is satisfiable if and only
if $\cE$ has a solution over $\N^*$.
\label{lma:necessaryInf}
\end{lemma}
\begin{proof}
If $\phi$ is satisfiable, then it has a model which is finite or
countably infinite. Now proceed as for Lemma~\ref{lma:necessary}.
For the converse, proceed as for Lemma~\ref{lma:sufficient}.
\end{proof}
\begin{lemma}
The set of constraints $\cE$ has a solution over $\N^*$ if and
only if it has a solution over $\{0, \aleph_0 \}$.
\label{lma:absorb}
\end{lemma}
\begin{proof}
Suppose $\cE$ has a solution $\theta: V \rightarrow \N^*$.  By
considering the forms in $\cE$, we see that $\theta': V \rightarrow
\{0, \aleph_0 \}$ defined by $\theta'(v) = \aleph_0 \theta'(v)$ is
also a solution. The other direction is trivial.
\end{proof}
Since the domain $\{0, \aleph_0 \}$ has only 2-elements, variables
interpreted over it are essentially Boolean.  If $x \in V$, let us
write $X$ for the corresponding statement $x = 0$, so that the
constraints $\cE$ are viewed as formulas of propositional logic. For
example, a constraint of the form
\begin{equation*}
x_1 + \cdots + x_n = x
\end{equation*}
becomes the set of Boolean formulas
\begin{equation*}
\{X_1 \wedge \cdots \wedge X_n \rightarrow X \} \cup
\{ X \rightarrow X_i \mid 1 \leq i \leq n \};
\end{equation*}
a constraint of the form
\begin{equation*}
x_1 + \cdots + x_n \geq 1
\end{equation*}
becomes the Boolean formula
\begin{equation*}
X_1 \wedge \cdots \wedge X_n \rightarrow \bot;
\end{equation*}
and a constraint of the form
\begin{equation*}
x > 0 \ \ \Rightarrow \ \ x_1 + \cdots + x_n \geq D
\end{equation*}
becomes the Boolean formula
\begin{equation*}
X_1 \wedge \cdots \wedge X_n \rightarrow X.
\end{equation*}

A quick check reveals that all of the resulting formulas
are Horn-clauses. This immediately yields:
\begin{theorem}[Kazakov]
The satisfiability problem for $\cGC^2$ is in EXPTIME.
\label{theo:kazakov}
\end{theorem}
The proof in Kazakov~\cite{logic:kazakov04} proceeds by showing that
satisfiability in $\cGC^2$ can be reduced in polynomial 
time to satisfiability in the 3-variable guarded fragment;
Theorem~\ref{theo:kazakov} then follows by the complexity bound for
the latter established by Gr\"{a}del~\cite{logic:gra99}. The approach
taken here is thus somewhat more direct. Moreover, Kazakov's reduction
is not conservative, and, as mentioned, yields no complexity bound for
the corresponding finite satisfiability problem.
\bibliographystyle{plain} \bibliography{logic}
\end{document}